# Experimental Investigation of the Hyperfine Structure of Tm I with Fourier Transform Spectroscopy Part B: in the NIR wavelength range from 700 nm to 2250 nm

Taha Yusuf Kebapcı[1,2], Sami Sert[1], Şeyma Parlatan[1,3], İpek Kanat Öztürk[4], Gönül Başar[4], Günay Başar[5], Maris Tamanis[6], Sophie Kröger[7]

[1]*Graduate School of Engineering and Sciences, Istanbul University, TR-34452 Beyazıt, Istanbul, Turkey*
[2]*Pamukkale University Kınıklı Campus, Denizli Health Services Vocational School of Higher Education, D Block TR-20070 Kınıklı, Denizli, Turkey*
[3]*Istinye University, Vocational School of Health Services,TR-34010,Zeytinburnu, Istanbul, Turkey*
[4]*Istanbul University, Faculty of Science, Physics Department, TR-34134 Vezneciler, Istanbul, Turkey*
[5]*Istanbul Technical University, Faculty of Science and Letters, Physics Engineering Department, TR-34469 Maslak, Istanbul, Turkey*
[6]*Laser Centre, University of Latvia, Jelgavas Street 3, LV-1004 Riga, Latvia*
[7]*Hochschule für Technik und Wirtschaft Berlin, Fachbereich 1, Wilhelminenhofstr. 75A, Berlin D-12459, Germany*

**Abstract**

In this study, we investigated the hyperfine structure of 43 spectral lines of atomic thulium. We analyzed Fourier-transform spectra in the wavelength range from 700 nm to 2250 nm, which corresponds to the wavenumber range from 14300 cm$^{-1}$ to 4440 cm$^{-1}$, respectively. The excited thulium atoms were generated in a hollow-cathode lamp. As a result of this investigation, the magnetic-dipole hyperfine constant *A* of 17 fine structure levels have been determined experimentally, 14 of them for the first time. The magnetic-dipole hyperfine constant values of the three remaining levels, reported in the literature, differed significantly from the results of our determination.

1. **Introduction**

Thulium (Tm, Z = 69) is the third to last element in the series of lanthanides. Due to the nuclear spin $I = \frac{1}{2}$ of the only stable isotope $^{169}$Tm, the hyperfine (hf) interaction leads to a splitting of a fine structure line into three or four hf peaks only, two strong peaks with high intensity, and one or two weak peaks with low intensity. If the weak peaks disappear in the noise, it is not possible to evaluate the hf structure of unresolved lines without additional information.

As described in detail in Part A [1], many studies of the spectrum of this element have already been carried out [2–28]. All experimental investigations of the hyperfine structure (hfs) of atomic Tm from the last decades were carried out with lasers and were limited to the visible or near infrared (NIR) range below 1000 nm [9–14, 16–18, 20–25, 27, 28]. Part A of this study [1] is also limited to the visible range. In the present work, the hfs of spectral lines in the NIR range up to 2250 nm is examined for the first time. Additionally, our advantage compared to earlier investigations of Fourier transformation (FT) spectra is that the results of laser spectroscopic investigations [2–5, 8], which have been carried out in the meantime, could be included in our evaluation. This enabled us to determine the magnetic-dipole hyperfine constant *A* for 14 energy levels for the first time and to obtain improved *A* values for three other levels.

## 2. Experiment

In the present experiments, a Tm plasma was generated in a hollow-cathode lamp with a current of about 70 mA. The hollow cathode was cooled with liquid nitrogen to reduce Doppler line broadening. Emission spectra have been recorded with a Bruker IFS 125HR Fourier-Transform spectrometer at the Laser Centre of the University of Latvia. For the range from 700 nm to 850 nm, a Hamamatsu R928 photomultiplier tube (PMT) was used as detector, whereas for the range from 850 nm to 2250 nm an InGaAs photodetector was used. The measurements were performed with an entrance aperture of 1.5 mm and spectral resolution of 0.03 cm$^{-1}$. In order to identify the spectral lines unambiguously as belonging to the element Tm, the spectra were recorded out twice: once with Ar and once with Ne as buffer gas in the hollow-cathode lamp. Each spectrum was obtained from the average of 50 scans. Only lines appearing in both spectra were assigned to the element Tm. Ar or Ne lines, were used to calibrate the spectra as accustomed.
Unfortunately, some ringing artifacts appear as spurious signals near some stronger lines, e.g., lines with a signal-to-noise ratio (SNR) of more than about 50, which somewhat limited the analysis.

## 3. Hyperfine-structure Analysis and Results

To determine magnetic dipole hfs constants, for selected lines a range of about 1 cm$^{-1}$ (~30 GHz), around the centre of gravity of the line is cut-out of the Tm-Ar-spectrum. The cut out parts then were analysed using the program Fitter [29]. The program Fitter performs an iterative least-squares fit of the experimental data using the hfs constants *A* of the upper and the lower levels, the center of gravity of the line, the line profile parameters, and the peak intensity parameters as fitting parameters. We chose Voigt profile functions for the hfs components with the same profile parameter for each component and coupled the peak intensity according to the theoretical intensity ratios for hfs transitions.

In our FT spectra, the linewidth results from the Doppler broadening in the hollow-cathode lamp as well as from the apparatus function of the FT spectrometer. The natural linewidth for the lines under investigation is significantly smaller and does not contribute substantially to the measured full width at half maximum (FWHM), which lies between 900 MHz and 1200 MHz for the investigated lines.

Initially, 40 lines were selected; three lines were added later. All lines are listed in Table 1. For our hfs analysis we chose lines for which the magnetic-dipole hfs constant *A* was unknown for one of the levels involved in the transition. When using Doppler-restricted measurement methods, in most of the Tm lines the two strong peaks are not resolved. The line selection was further narrowed by choosing only lines with well-resolved strong hfs peaks. Even for these lines, the weak peaks are almost always not clearly recognizable, i.e., they disappear under the large peaks as shown in Figure 1a, or they disappear in the noise or ringing artifacts as shown in Figure 1b. Thus, it was not possible to determine both hfs constants *A* of the lower and of the upper energy levels of a spectral line. Therefore, we evaluated only those lines for which the hfs constant for one of the two levels was already known from the literature. This hfs constant was fixed during the fitting procedure. The list of fixed values is given in Table 2. In the 40 selected lines, 15 different levels with unknown hfs constant *A* are involved.

By analysing the entire spectrum and looking for lines with two well-resolved peaks, one line caught our eye, because both A constants had been reported as known, but the structure did not fit the known values. This line was located at 1863.263 nm or 5365.464 cm$^{-1}$, and classified as a transition from 37724.84 cm$^{-1}$, $J = 5/2$ with even parity to 32359.372 cm$^{-1}$, $J = 7/2$. For the upper level, the value $A = -2263.3$ (3.9) MHz is given in [15] and stated to be obtained by using only one line. The hfs constant $A = -612.9$ (1.2) MHz of the lower energy level of this line comes from [18] and was measured and confirmed with five different lines. A simulation with these hfs constants is shown in Figure 2a. The starting position of the strong component was placed at the position of the strongest component of the experimental spectrum. As can be seen clearly, this structure does not fit to the observed pattern. We suggest that the problem is related to the *A* constant of the upper level, which is determined by using only one line. Therefore, we searched for more lines in which the same upper level is involved. We found a line at 1905.873 nm or 5245.507 cm$^{-1}$, classified as a transition from 37724.84 cm$^{-1}$, $J = 5/2$ with even parity to 32479.326 cm$^{-1}$, $J = 5/2$. Again, the *A* constant of the lower level is known from the literature (see Table 2). A simulation with the literature values for the hfs constants also here shows a strong deviation, as shown in Figure 2c. If the constant of the upper levels is now fitted as a free parameter, both lines lead to a similar $A = -15.7(1.8)$ MHz and $A = -24(13)$ MHz, respectively (see Table 3). These two values agree within their error limits, but differ greatly from the value quoted in the literature. The best-fitted spectra describe line profile well in both cases, as can be seen in Figures 2b and 2d. Therefore, we proved that the literature value given in [15] for the level is not correct and we present a corrected value in Table 3.

For two other levels, namely the ones at 24246.425 cm$^{-1}$, $J = 7/2$ with the odd parity and 25745.117 cm$^{-1}$, $J = 5/2$ with the even parity there were also discrepancies of a similar kind. In many other cases, the deviations are slightly higher than the range of the error bars, so it is not easy to decide which of the *A* values is more reliable.

The level at 25745.117 cm$^{-1}$ was involved in several of the transitions that we chose for determining other constants. In each case, it was the value associated with that level's fixed *A* value that was out of bounds compared with the results from other lines. Thus, we looked for additional lines that involve this level and were finally able to determine a revised *A* constant (see Table 4). With this revised *A* value, on the other hand, a much better agreement could be found for the new *A* constants of other levels.

The level at 24246.425 cm$^{-1}$ has already been noticed as problematic in part A of this study [1]. Also in the present work, for several lines in which this level is involved the results did not fit well. Here, too, a revised value could be found with the appropriate lines (see Table 3). With this revised *A* value, some lines of part A [1] were re-analyzed afterward and now show good agreement with our revised A values.

**Table 1:** Lines in the FT spectrum which were analysed to determine the magnetic-dipole hyperfine-structure constants *A*.

| Line | | | | upper level | | | lower level | | |
|---|---|---|---|---|---|---|---|---|---|
| $\lambda_{air}$ /nm | int. | SNR | $\sigma$ / cm$^{-1}$ | $E_{up}$/cm$^{-1}$ | $P_{up}$ | $J_{up}$ | $E_{low}$/cm$^{-1}$ | $P_{low}$ | $J_{low}$ |
| 721.4173 | 100 | 14 | 13857.78 | 39602.902 | e | 5/2 | 25745.117 | o | 3/2 |
| 728.3777 | 60 | 19 | 13725.36 | 39470.489 | e | 5/2 | 25745.117 | o | 5/2 |
| 738.1009 | 10 | 8 | 13544.55 | 38502.00 | e | 9/2 | 24957.469 | o | 11/2 |
| 738.7883 | 100 | 16 | 13531.95 | 39277.087 | e | 5/2 | 25745.117 | o | 5/2 |
| 770.1423 | 400 | 53 | 12981.04 | 38502.00 | e | 9/2 | 25520.987 | o | 11/2 |
| 1017.307 | 200 | 240 | 9827.181 | 27440.858 | o | 9/2 | 17613.659 | e | 9/2 |
| 1078.464 | 4 | 15 | 9269.907 | 31007.60 | o | 11/2 | 21737.685 | e | 9/2 |
| 1109.560 | 20 | 63 | 9010.114 | 31007.60 | o | 11/2 | 21997.473 | e | 11/2 |
| 1162.005 | - | 70 | 8603.460 | 27440.858 | o | 9/2 | 18837.385 | e | 9/2 |
| 1183.048 | - | 52 | 8450.429 | 27440.858 | o | 9/2 | 18990.406 | e | 11/2 |
| 1226.886 | - | 39 | 8148.487 | 34587.982 | o | 9/2 | 26439.491 | e | 7/2 |
| 1255.021 | - | 234 | 7965.815 | 24708.041 | o | 7/2 | 16742.237 | e | 7/2 |
| 1270.451 | - | 282 | 7869.068 | 24611.303 | o | 5/2 | 16742.237 | e | 7/2 |
| 1306.098 | - | 211 | 7654.300 | 24611.303 | o | 5/2 | 16957.006 | e | 7/2 |
| 1309.105 | - | 819 | 7636.718 | 33292.78 | o | 5/2 | 25656.019 | e | 5/2 |
| 1318.152 | - | 73 | 7584.304 | 33240.362 | o | 7/2 | 25656.019 | e | 5/2 |
| 1332.227 | - | 4257 | 7504.176 | 24246.425 | o | 7/2 | 16742.237 | e | 7/2 |
| 1333.822 | - | 24 | 7495.202 | 33240.362 | o | 7/2 | 25745.117 | e | 5/2 |
| 1358.752 | - | 195 | 7357.683 | 24701.058 | o | 9/2 | 17343.374 | e | 7/2 |
| 1371.479 | - | 2125 | 7289.405 | 24246.425 | o | 7/2 | 16957.006 | e | 7/2 |
| 1409.179 | - | 525 | 7094.391 | 24708.041 | o | 7/2 | 17613.659 | e | 9/2 |
| 1437.117 | - | 419 | 6956.474 | 33395.984 | o | 7/2 | 26439.491 | e | 7/2 |
| 1458.759 | - | 33 | 6853.269 | 33292.78 | o | 5/2 | 26439.491 | e | 7/2 |
| 1485.552 | - | 29 | 6729.665 | 32856.613 | o | 5/2 | 26126.907 | e | 5/2 |
| 1501.342 | - | 111 | 6658.888 | 31007.60 | o | 11/2 | 24348.692 | e | 9/2 |
| 1507.259 | - | 1958 | 6632.747 | 24246.425 | o | 7/2 | 17613.659 | e | 9/2 |
| 1528.872 | - | 44 | 6538.983 | 33240.362 | o | 7/2 | 26701.325 | e | 7/2 |
| 1557.911 | - | 102 | 6417.098 | 32856.613 | o | 5/2 | 26439.491 | e | 7/2 |
| 1634.155 | - | 149 | 6117.699 | 31773.78 | o | 5/2 | 25656.019 | e | 5/2 |
| 1657.318 | - | 43 | 6032.197 | 34587.982 | o | 9/2 | 28555.799 | e | 7/2 |
| 1658.307 | - | 27 | 6028.600 | 31773.78 | o | 5/2 | 25745.117 | e | 5/2 |
| 1760.853 | - | 3072 | 5677.515 | 22419.764 | o | 9/2 | 16742.237 | e | 7/2 |
| 1830.082 | - | 253 | 5462.744 | 22419.764 | o | 9/2 | 16957.006 | e | 7/2 |
| 1863.263 | - | 135 | 5365.464 | 37724.84 | e | 5/2 | 32359.372 | o | 7/2 |
| 1898.626 | - | 477 | 5265.529 | 30921.58 | o | 7/2 | 25656.019 | e | 5/2 |
| 1905.873 | - | 16 | 5245.507 | 37724.84 | e | 5/2 | 32479.326 | o | 5/2 |
| 2015.781 | - | 59 | 4959.502 | 24708.041 | o | 7/2 | 19748.543 | e | 9/2 |
| 2018.626 | - | 1254 | 4952.513 | 24701.058 | o | 9/2 | 19748.543 | e | 9/2 |
| 2065.473 | - | 20 | 4840.185 | 33395.984 | o | 7/2 | 28555.799 | e | 7/2 |
| 2080.126 | - | 723 | 4806.089 | 22419.764 | o | 9/2 | 17613.659 | e | 9/2 |
| 2121.243 | - | 27 | 4712.931 | 32856.613 | o | 5/2 | 28143.67 | e | 3/2 |
| 2222.674 | - | 79 | 4497.858 | 24246.425 | o | 7/2 | 19748.543 | e | 9/2 |
| 2230.501 | - | 17 | 4482.074 | 30921.58 | o | 7/2 | 26439.491 | e | 7/2 |

**Note:** $\sigma$: wavenumber of the center of gravity resulting from the hfs fit; $\lambda_{air}$: wavelengths in air, calculated from the $\sigma$ values according to [30]; int: intensity according to [8]; SNR: signal-to-noise ratio in the Tm-Ar spectrum; $E_{up}$, $E_{low}$, $p_{up}$, $p_{low}$, $J_{up}$, and $J_{low}$: level energy, parity and total electron angular momentum of the upper and lower level, respectively.

**Table 2:** Experimental magnetic-dipole hfs constants $A$ of Tm I from the literature that were fixed during the fit.

| Energy /cm$^{-1}$ | $J$ | Parity | $A$ /MHz | Ref. |
|---|---|---|---|---|
| 16742.237 | 7/2 | e | -736.6(1.0) | [11] |
| 16957.006 | 7/2 | e | -491.1(1.0) | [11] |
| 17343.374 | 7/2 | e | -166.24(8) | [9] |
| 17454.818 | 13/2 | e | -365.91(55) | [13] |
| 17613.659 | 9/2 | e | -629.25(8) | [9] |
| 18837.385 | 9/2 | e | -422.4(9) | [4] * |
| 18990.406 | 11/2 | e | -581.4(1.3) | [17] |
| 19748.543 | 9/2 | e | -694.8(4) | [17] |
| 19753.830 | 7/2 | e | -536.6(9) | [5] * |
| 21737.685 | 9/2 | e | -357.1(1.5) | [5]* |
| 21997.473 | 11/2 | e | -342.8(1.3) | [21] |
| 24246.425 | 7/2 | o | -80.9(1.3) | t.w. |
| 24348.692 | 9/2 | e | -371.7(1.5) | [5] * |
| 24418.018 | 5/2 | e | -659(2) | [5] * |
| 24708.041 | 7/2 | o | -207.5(1.2) | t.w. |
| 24957.469 | 11/2 | o | -538.7(1.5) | [21] |
| 25520.987 | 11/2 | o | -234.9(8) | [21] |
| 25656.019 | 5/2 | e | -410.7(1.4) | [11] |
| 25745.117 | 5/2 | e | -593(5) | t.w.† |
| 26126.907 | 5/2 | e | -1161.47(1.20) | [20] |
| 26439.491 | 7/2 | e | -908.42(30) | [9] |
| 26701.325 | 7/2 | e | -426(15) | [26] |
| 27440.858 | 9/2 | o | -166.8(1.6) | t.w. |
| 28143.67 | 3/2 | e | -475.6(8) | [18] |
| 28555.799 | 7/2 | e | -992.4(1.5) | [18] |
| 32359.372 | 7/2 | o | -612.9(4) | [18] |
| 32479.326 | 5/2 | o | 383.0(7) | [18] |
| 39277.087 | 5/2 | o | -1280.1(2.0) | [1] |
| 39470.489 | 5/2 | o | 175(5) | [1] |
| 39602.902 | 3/2 | o | -1911(9) | [1] |

t.w.: this work
* Converted from 10$^{-3}$ cm$^{-1}$ scale.
† revised value, see Table 3

**Table 3:** New magnetic-dipole hyperfine structure constant, A for levels of even parity of Tm I.

| $E$ / cm$^{-1}$ | $J$ | $\sigma$ /cm$^{-1}$ | $A_{exp}$ / MHz | $A_{mean}$ / MHz | $A_{calc}$ / MHz [17] |
|---|---|---|---|---|---|
| 25745.117 | 5/2 | 13531.959 | -593 (7) | | |
| | | 13725.359 | -594 (7) | -593 (5)* | -475 |
| | | 13857.783 | -586 (15) | | |
| 37724.84 | 5/2 | 5245.507 | -24 (13) | -15.9 (1.8)** | -420 |
| | | 5365.464 | -15.7 (1.8) | | |
| 38502.00 | 9/2 | 13544.552 | 36 (5) | 41.0 (1.6) | -298 |
| | | 12981.041 | 41.5 (1.6) | | |

* revised value, in [20] given with $A = -439.38(1.00)$ MHz
** revised value, in [15] given with $A = -2263.3\,(3.9)$ MHz

**Table 4:** New magnetic-dipole hyperfine-structure constant, A for levels of odd parity of Tm I.

| $E$ / cm$^{-1}$ | $J$ | $\sigma$ /cm$^{-1}$ | $A_{exp}$ / MHz | $A_{mean}$ / MHz | $A_{calc}$ / MHz [17] |
|---|---|---|---|---|---|
| 22419.764 | 9/2 | 5677.51515 | 41.1 (1.8) | 40.8 (1.0) | 43 |
| | | 5462.74427 | 40.4 (1.5) | | |
| | | 4806.08915 | 41.0 (1.8) | | |
| 24246.425 | 7/2 | 7504.17599 | -81.2 (2.3) | -80.9 (1.2)* | -94 |
| | | 7289.40535 | -80.0 (2.2) | | |
| | | 6632.74711 | -81.7 (2.2) | | |
| | | 4497.85773 | -80.4 (2.7) | | |
| 24611.303 | 5/2 | 7869.06776 | -9.3 (2.8) | -9.9 (2.0) | -68 |
| | | 7654.29969 | -10.4 (2.6) | | |
| 24701.058 | 9/2 | 7357.68279 | -446.7 (1.9) | -441.7 (1.5) | -440 |
| | | 4952.51264 | -440.6 (0.9) | | |
| 24708.041 | 7/2 | 7965.81473 | -207.9 (2.7) | -207.5 (1.2) | -186 |
| | | 7094.39099 | -207.9 (1.5) | | |
| | | 4959.50243 | -206.2 (2.5) | | |
| 27440.858 | 9/2 | 9827.1806 | -169.0 (1.4) | -166.8 (1.6) | -179 |
| | | 8603.45964 | -164.0 (1.5) | | |
| | | 8450.42934 | -167.1 (1.9) | | |
| 30921.58 | 7/2 | 5265.52901 | -81.5 (2.7) | -82.4 (2.2) | -70 |
| | | 4482.07441 | -84 (3) | | |
| 31007.60 | 11/2 | 9269.90727 | -530 (6) | -534.3 (1.3) | -530 |
| | | 9010.11393 | -534.4 (2.0) | | |
| | | 6658.88766 | -534.5 (1.8) | | |
| 31773.78 | 5/2 | 6117.69923 | -1241 (3) | -1243 (3) | -1248 |
| | | 6028.59962 | -1255 (8) | | |
| 32856.613 | 5/2 | 6729.66518 | -702 (5) | -701.9 (2.6) | -756 |
| | | 6417.09841 | -700.2 (2.0) | | |
| | | 4712.93064 | -718 (6) | | |
| 33240.362 | 7/2 | 7584.30415 | -922.6 (2.1) | -923.1 (2.1) | -913 |
| | | 7495.20243 | -929 (7) | | |
| | | 6538.98297 | -921 (15) | | |
| 33292.78 | 5/2 | 7636.71788 | 25.9 (2.7) | 25.7 (2.3) | 5 |
| | | 6853.26864 | 25 (5) | | |
| 33395.984 | 7/2 | 6956.47403 | -554.5 (1.6) | -555.4 (1.9) | -616 |
| | | 4840.18476 | -564 (5) | | |
| 34587.982 | 9/2 | 8148.48674 | -433.3 (2.1) | -434.4 (2.4) | -446 |
| | | 6032.19717 | -442 (6) | | |

* revised value, in [21] given with $A = -197(5)$ MHz

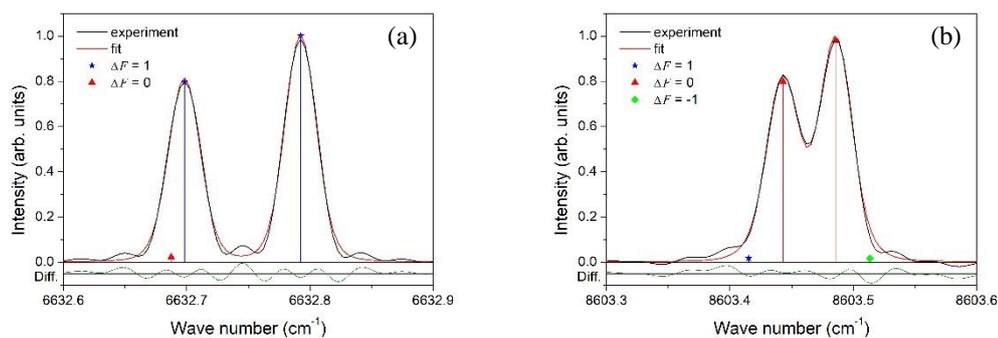

**Figure 1:** Examples for experimental spectrum together with corresponding best fit. In the lower part, the residual is given (difference between experimental and fitted spectrum.
  a) transition 24246.425 cm$^{-1}$ ($J = 7/2$) $\to$ 17613.659 cm$^{-1}$ ($J = 9/2$) at $\sigma = 6632.747$ cm$^{-1}$.
  b) transition 27440.858 cm$^{-1}$ ($J = 9/2$) $\to$ 18837.385 cm$^{-1}$ ($J = 9/2$) at $\sigma = 8603.462$ cm$^{-1}$

The resulting magnetic-dipole hyperfine-structure constants $A$ are listed in Tables 3 and 4 for the levels of even and odd parity, respectively. The wave number given in the third column specifies the line that was used to determine the hyperfine constant $A_{exp}$. In the penultimate column, the weighted mean value $A_{mean}$ is given, calculated using the error bars of $A_{exp}$ as weights. For the calculation of the error bars, the same procedure is used as explained in Part A [1].

In the last column the values calculated from a semi-empiric analysis of the fine-structure and hfs constants from the literature [17] are given for comparison. For even parity levels, the deviation between the experimental value $A_{mean}$ and the calculated value $A_{calc}$ is very large. The $A$ values for the odd parity levels mostly agree well.

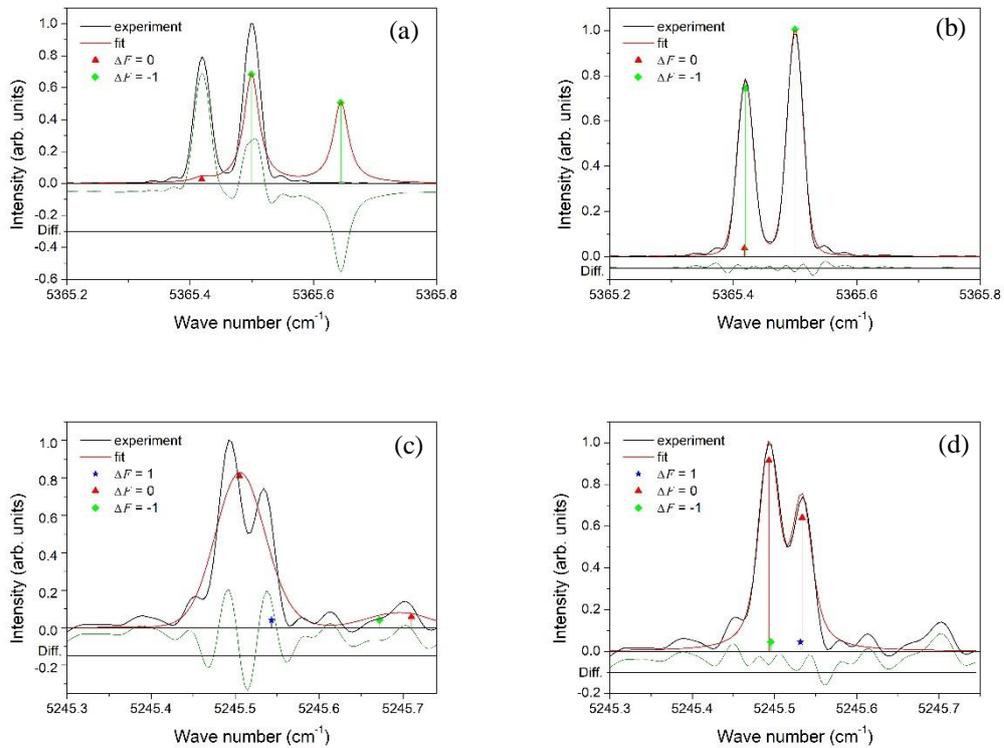

**Figure 2:** (a) The spectrum of the best-fit result obtained by taking the $A$ value from the literature; (b) The spectrum of the best-fit result obtained as a result of the fit process using the value $A$ re-determined in this study; the best-fit spectra obtained by the fit process with the (c) incorrect and (d) revised $A$ values.

a) transition 37724.84 cm$^{-1}$ ($J$=5/2) → 32359.372 cm$^{-1}$ ($J$=7/2) at $\sigma$ = 5365.46 cm$^{-1}$
b) transition 37724.84 cm$^{-1}$ ($J$=5/2) → 32479.326 cm$^{-1}$ ($J$=5/2) at $\sigma$ = 5245.49 cm$^{-1}$

Although it is not the subject of present work, we also looked at the deviations of the experimental centre of gravity wavenumbers compared with the differences of level-energy values. Quite large deviations were observed for some lines. A closer inspection reveals that the greatest deviations belong to transitions that are connected to the same upper levels. Two levels are particularly badly affected: the ones at 31773.78 cm$^{-1}$ and 33240.362 cm$^{-1}$. For the lines connected with these levels, the distance between the two main hfs components is so great that the two peaks appear to be separate lines, but in fact, it is only one line with a very broad hfs. We assume, that only one of the two hfs peaks of the transition was taken into account when determining the level energies. For all these lines, the distance between the two hfs peaks is between 0.04 cm$^{-1}$ and 0.06 cm$^{-1}$, which corresponds to magnitude of the deviations in the level energies.

4. **Conclusion and Outlook**

The hyperfine structure of 43 selected lines in high-resolution FT spectra of Tm-Ar plasma has been analyzed. Each of these lines have two strong, separated hyperfine components. By analysing these 43 spectral lines, we determined the magnetic-dipole hfs constants *A* for 17 levels, of which 3 had even parity and 14 had odd parity. For 14 levels, one of even parity and 13 of odd parity, the hfs constants have been measured for the first time. The values of three hfs constants, which had been reported in the literature previously, have been corrected.

In addition to the lines, examined in this study, plenty of unresolved Tm lines can be seen in the FT spectra. The analysis of these lines poses an even greater challenge, since a single line does not provide enough information to determine the mean values for the hfs constants. Therefore, many lines should be evaluated in a common fit. This task is planned for the next study.

A repetition of the semi-empiric calculation of the fine structure and hfs, based on an expanded database including the present new experimental data, is recommended.

**Acknowledgements**

We acknowledge Ruvin Ferber and Artis Kruzins from the Laser Centre of the University of Latvia for assistance in the experiment. This work has been supported by the Istanbul University Scientific Research Project via project No. 33657 and also funded by Scientific Research Projects Coordination Unit of Istanbul University Project No.: 30048, as well as by Latvian Council of Science, project No. lzp-2020/1-088: "Advanced spectroscopic methods and tools for the study of evolved stars".